\documentclass{aa}  

\usepackage{graphicx}
\usepackage{txfonts}

\begin{document}

\title{Determination of the relativistic parameter $\gamma$ using 
very long baseline interferometry}

\author{S. B. Lambert \and C. Le Poncin-Lafitte}

\offprints{S. B. Lambert, \email{sebastien.lambert@obspm.fr}}

\institute{Observatoire de Paris, D\'epartement Syst\`emes de R\'ef\'erence Temps Espace (SYRTE), CNRS/UMR8630, 75014 Paris, France}

\date{}

\titlerunning{Determination of $\gamma$ using VLBI}

\abstract
{}
{Relativistic bending in the vicinity of a massive body is characterized only 
by the post-Newtonian parameter $\gamma$ within the standard parameterized 
post-Newtonian formalism, which is unity in General Relativity. Aiming at 
estimating this parameter, we use very long baseline interferometry (VLBI) to 
measure the gravitational deflection of radio waves emitted by distant compact 
radio sources, by Solar System bodies.}
{We analyze geodetic VLBI observations recorded since 1979. We compare 
estimates of $\gamma$ and errors obtained using various analysis schemes 
including global estimations over several time spans and with various 
Sun elongation cut-off angles, and analysis of radio source coordinate 
time series.}
{We arrive at the conclusion that the relativistic parameter
$\gamma$ cannot be estimated at better than $2\times10^{-4}$. The
main factor of limitation is the uncertainty in the determination of 
(global or session-wise) radio source coordinates. A sum of various
instrumental and modeling errors and analysis strategy defects, that cannot be
decorrelated and corrected yet, is at the origin of the limitating noise.}
{}
\keywords{Astrometry -- Relativity -- Techniques: interferometric}

\maketitle

\section{Introduction}

One of the cornerstones of test of General Relativity (GR) is the measurement 
of light deflection in the vicinity of the Sun. In the parameterized 
post-Newtonian (PPN) formalism (Will~1993), which contains 10 parameters, the 
predicted angle of deflection $\theta$ is
\begin{equation}
\theta\approx(\gamma+1)\frac{GM}{c^2b}(1+\cos\phi), 
\end{equation}
where $G$ is the Newtonian gravitational constant, $c$ the speed of light in a 
vacuum, $M$ the mass of the deflecting body, $b$ the impact parameter
{\bf (defined as the minimal distance of the ray to the center 
of mass of the deflecting body)}, $\phi$ 
the elongation angle between the deflecting body and the source as viewed by 
the observer and $\gamma$ is the PPN parameter characterizing the 
space curvature due to gravity. (See, e.g., Misner et al.~1973, 
Will~1993, and more generally speaking for an axisymmetric body, Le 
Poncin-Lafitte \& Teyssandier~2008.) Thus, a grazing ray at the Sun's limb 
is deflected by $\sim$1.7$^{\prime\prime}$. In GR $\gamma=1$. It is crucial to 
note that light deflection experiments give us privileged access to $\gamma$, 
independently from other post-Newtonian parameters. This point is even more 
important when one thinks that cosmological models (Damour \& Polyakov~1994,
Damour et al.~2002) predict deviations of $|\gamma-1|$ of the order of 
$10^{-6}-10^{-7}$.

\begin{figure}[htbp]
\begin{center}
\includegraphics[width=8.5cm]{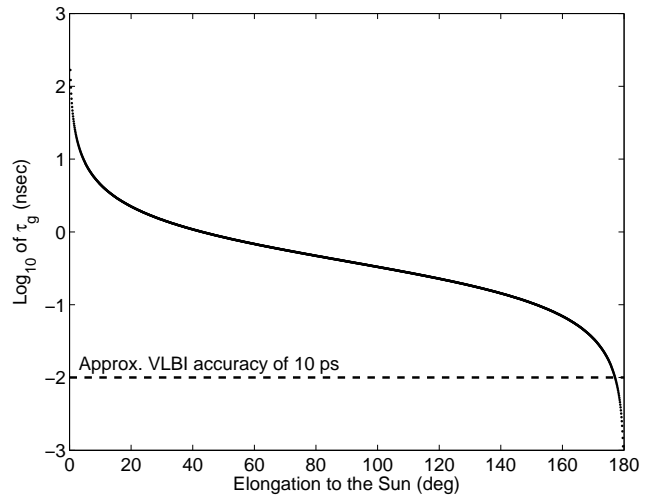}
\end{center}
\caption{Gravitational delay $\tau_g$ as a function of the 
elongation angle $\phi$ to the Sun for the baseline Westford--Wettzell.}
\label{figww}
\end{figure}

{\bf Very long baseline radio interferometry (VLBI) is sensitive to space-time 
curvature through the gravitational time delay, given by (e.g, Finkelstein 
et al.~1983)}
\begin{equation}
\tau_g=(\gamma+1)\frac{GM}{c^3}{\rm log}\left(\frac{|\vec r_1|+\vec r_1.\vec k}{|\vec r_2|+\vec r_2.\vec k}\right),
\end{equation}
{\bf where $\vec r_i$ stands for the position vector of the 
$i$th station and $\vec k$ the unit vector pointing towards the radio source,
both referred to the center of mass of the deflecting body. For a typical
VLBI baseline between Westford (Massachusetts) and Wettzell (Germany) of
$\sim$6,000~km, $\tau_g$ is $\sim$170~nanoseconds (ns) for a source 
at the Sun's limb, rapidly decreases to $\sim$10~ns at 4$^{\circ}$ 
away from the Sun, and 
remains of the order of the accuracy of VLBI measurements 
(nowadays around 10~ps) even for elongations close to 180$^\circ$ 
(see Fig.~\ref{figww}).}

VLBI is used on a regular basis since the early 1980's for monitoring 
Earth orientation and estimating station 
displacements and extragalactic radio source coordinates at 2 and 8~GHz. 
The number of {\bf radio sources} per session as well as the 
data recording reliability
have drastically improved in the last decade. About 4,000 diurnal
session files, representing more than 5~millions delays, are made 
available through the International VLBI Service for Geodesy and
Astrometry (IVS, Schl\"uter \& Behrend~2007) data base.

{\bf The large amount of data from the permanent geodetic VLBI program 
can provide a number of tests of GR (Soffel et al.~1986). In the past 
years, VLBI data were used in various attempts to determine $\gamma$.
Robertson \& Carter~(1984), using less than 4 years of observations, 
found $\gamma$ consistent with GR within 0.005.} Robertson et 
al.~(1991), using 10 years of observations, estimated a standard error of 
0.002. Lebach et al.~(1995) got $0.9996\pm0.0017$ after observations of the 
relative deflection of 3C~273B and 3C~279. Shapiro et al.~(2004) obtained
$0.99983\pm0.00026$ (statistical standard error) using VLBI observations 
before~1999. The current best estimate of $\gamma$, however, 
was not obtained with VLBI: it is consistent with GR with an error 
of~$2\times 10^{-5}$, and was obtained by Bertotti et al.~(2005) who 
derived it from spacecraft tracking experiments.

{\bf Note that errors reported in the various papers are often 
formal errors obtained from the propagation through the adjustment
procedure of an initial SNR-derived standard error on the delays. 
They might therefore not directly compare to one another.

Though the above-referenced works (except Bertotti et al.) dealt with
deflection of the radio waves by the Sun, it must be mentioned that 
special VLBI sessions were carried out to measure the deflection 
close to Jupiter or other planets (Schuh et al.~1988).}

In this work, we estimate $\gamma$ from routine geodetic VLBI 
observations, using the additional 1999--2008 time period with respect 
to Shapiro et al. We compare estimates and errors obtained over 
several time spans and using various analysis schemes in order to
address the accuracy, and to point out some systematics and limitations.

\section{Close approaches to the Sun}

\begin{figure}[htbp]
\begin{center}
\includegraphics[width=8.5cm]{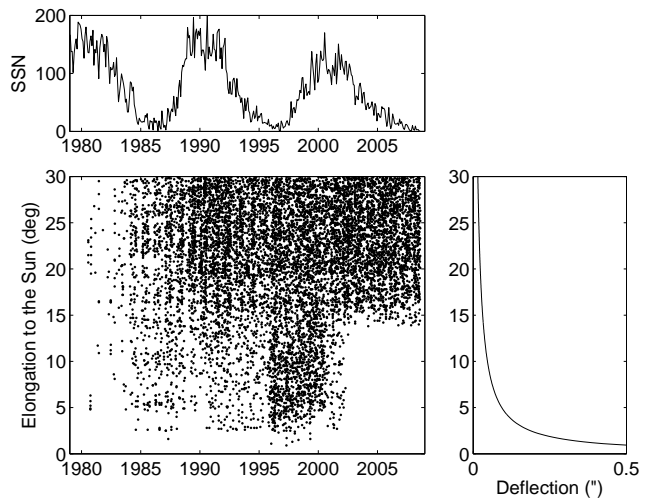}
\end{center}
\caption{Observational history of the sources at less than 30$^{\circ}$ 
to the Sun. The upper plot gives the Sun spot number (SSN, 
Clette et al.~2007). The right plot displays the deflection angle 
as predicted by~GR.}
\label{figelon}
\end{figure}

A set of 3,937 24-hr geodetic VLBI sessions, consisting of about 4.5 
million delays, will be fully or partly processed in the upcoming 
analyses. During the period that covers 3 August 1979--28
August 2008, the VLBI observing schedule included a number of radio 
sources that were observed at less than 15$^{\circ}$ to the sun. As 
it shows up in Fig.~\ref{figelon}, this number was weak before 1984, 
quite uniform during 1984--1996. Then it substantially increased 
during 1996--2002. It is worth noting that 1992--1999, that contains a 
number of close approaches, is a period of low solar activity. Since 
2002, the scheduling software at the IVS coordinating center was set 
with a minimal distance to the Sun at 15$^{\circ}$.
Fig.~\ref{figelon} naturally yields several time spans on which 
the analyses can be done: 1979--2008, which is the maximum 
number of available data, 1984--2008, that drops the early 
VLBI network, 1996--2002, which 
shows the highest density of close approaches, and 1984--2002, which 
represents a compromise between a high density of close approaches 
and a large number of data. Additionally, we also consider 1979--1999, 
as done in Shapiro et al., in order to check that we are consistent 
with their results. Finally, we would like to address two time 
spans that cover periods of low and high solar activity.
It is nevertheless difficult to keep the same characteristics (number of
sessions, number of sources, density of close approaches) for these
two periods since the VLBI observing program undergoes a continuous evolution.
We propose the three following time spans: 1994--1997, 
1998--2002a (starts 01/1998, and has approximately the same number of 
sessions and sources than 1994--1997), and 1998--2002b (starts 07/1998, and 
has approximately the same number of {\bf delays as 1994--1997}).

All our VLBI delays have been corrected from delay due to the
radio wave crossing of dispersive region in the signal propagation
path in a preliminary step that made use of 2~GHz and 8~GHz recordings.
Then, we only use the 8~GHz delays to fit the parameters listed in the 
next section. In the case of targets that are close to the Sun, the
relevant dispersive regions are the Earth's ionosphere and the solar
coronal plasma. Although approximated, the model for plasma delay 
correction as a function of electronic content and frequency should 
lead to errors of a few picoseconds, following Lebach et al.~(1995). 
(The authors mentioned this magnitude for a period of low solar 
activity. During periods of higher activity, the electronic content 
can be several times higher.) The reader must therefore keep in mind this 
order of magnitude when potential sources of limitation will be 
listed in further sections. Additionally, an error in the solar coronal 
plasma delay correction would lead to a falsified estimate of $\gamma$,
since the plasma-induced deflection would be absorbed there in.
Rather than a relativistic parameter, $\gamma$ would therefore be 
simply considered as a ``deflection" parameter.

\begin{table*}[htbp]
\caption[]{Characteristics of the solutions and estimates of $\gamma$.}
\label{tab01}
\begin{center}
\begin{tabular}{lrrrcrr}
\hline
\hline
\noalign{\smallskip}
 & No. sessions & No. delays & No. sources & Postfit rms delay & $\gamma$ & $\chi^2/f$ \\
 & & & & (ps) & & \\
\noalign{\smallskip}
\hline
\noalign{\smallskip}
1979--2008 & 3,937 & 4,386,112 & 988 & 25.0 & $0.99984 \pm 0.00015$ & 0.86 \\
\noalign{\smallskip}
1984--2008 & 3,852 & 4,348,913 & 988 & 24.9 & $0.99986 \pm 0.00015$ & 0.86 \\
\noalign{\smallskip}
1984--2002 & 3,040 & 2,857,624 & 781 & 27.0 & $0.99993 \pm 0.00017$ & 0.89 \\
\noalign{\smallskip}
1979--1999 & 2,598 & 2,115,509 & 723 & 27.4 & $0.99983 \pm 0.00020$ & 0.91 \\
\noalign{\smallskip}
1996--2002 &   753 & 1,024,322 & 676 & 27.5 & $0.99940 \pm 0.00022$ & 0.83 \\
\noalign{\smallskip}
1994--1997 &   650 &   849,084 & 683 & 24.6 & $0.99968 \pm 0.00024$ & 0.83 \\
\noalign{\smallskip}
1998--2002a &  650 &   953,882 & 643 & 26.3 & $1.00017 \pm 0.00032$ & 0.81 \\
\noalign{\smallskip}
1998--2002b &  595 &   873,827 & 616 & 26.2 & $1.00031 \pm 0.00035$ & 0.82 \\
\noalign{\smallskip}
\hline
\end{tabular}
\end{center}
\end{table*}

\section{Data analysis and results}

\subsection{Global solutions}\label{secglo}

We run global solutions over the aforementioned time spans. 
In all these solutions, the Earth orientation parameters
and the station coordinates are estimated once per session.
$\gamma$ is estimated as a global parameter. Source coordinates are also
estimated as global parameters without global constraint: the sources 
are allowed to stay within circles of $10^{-8}$~rad diameter 
around a priori positions. The choice of the a priori
catalogue for source coordinates is discussed later.

Now, we quickly go on with some technical characteristics of the solutions. 
The cut-off elevation angle is set to 5$^{\circ}$. {\bf A priori zenith 
delays are determined from local pressure values (Saastamoinen 1972) 
which are then mapped to the elevation of the observation using the 
Niell mapping function (Niell 1996). Zenith wet delays are estimated 
as a continuous piece-wise linear function at 20-min interval.} 
Troposphere gradients are estimated as 8-hr East and North piece-wise 
functions at all stations except a set of 110 stations having poor 
observational history. Station heights are corrected from atmospheric 
pressure and oceanic tidal loading. The relevant loading quantities 
are deduced from surface pressure grids from the U.~S. NCEP/NCAR 
reanalysis project atmospheric global circulation model (Kalnay et al.~1996)
and from the GOT00.2 ocean tide model (Ray~1999, Petrov \& Boy~2004). 
No-net rotation constraint per session is applied to the positions of 
all stations, excluding HRAS~085 (Fort Davis, Texas) and Fairbanks (Alaska) 
because of strong non-linear displacements. (The latter site undergoes 
post-seismic relaxation effects after a large earthquake on Denali fault 
in 2003. See, e.g., MacMillan \& Cohen~2004, Titov \& Tregoning 2004, 2005.)
All the calculations use the Calc~10.0/Solve~2006.06.08 geodetic VLBI 
analysis software package and are {\bf carried out} at the Paris 
Observatory IVS Analysis Center (Gontier et al.~2008). Results are 
reported in Table~\ref{tab01}.

\begin{figure*}[htbp]
\begin{center}
\includegraphics[width=8cm]{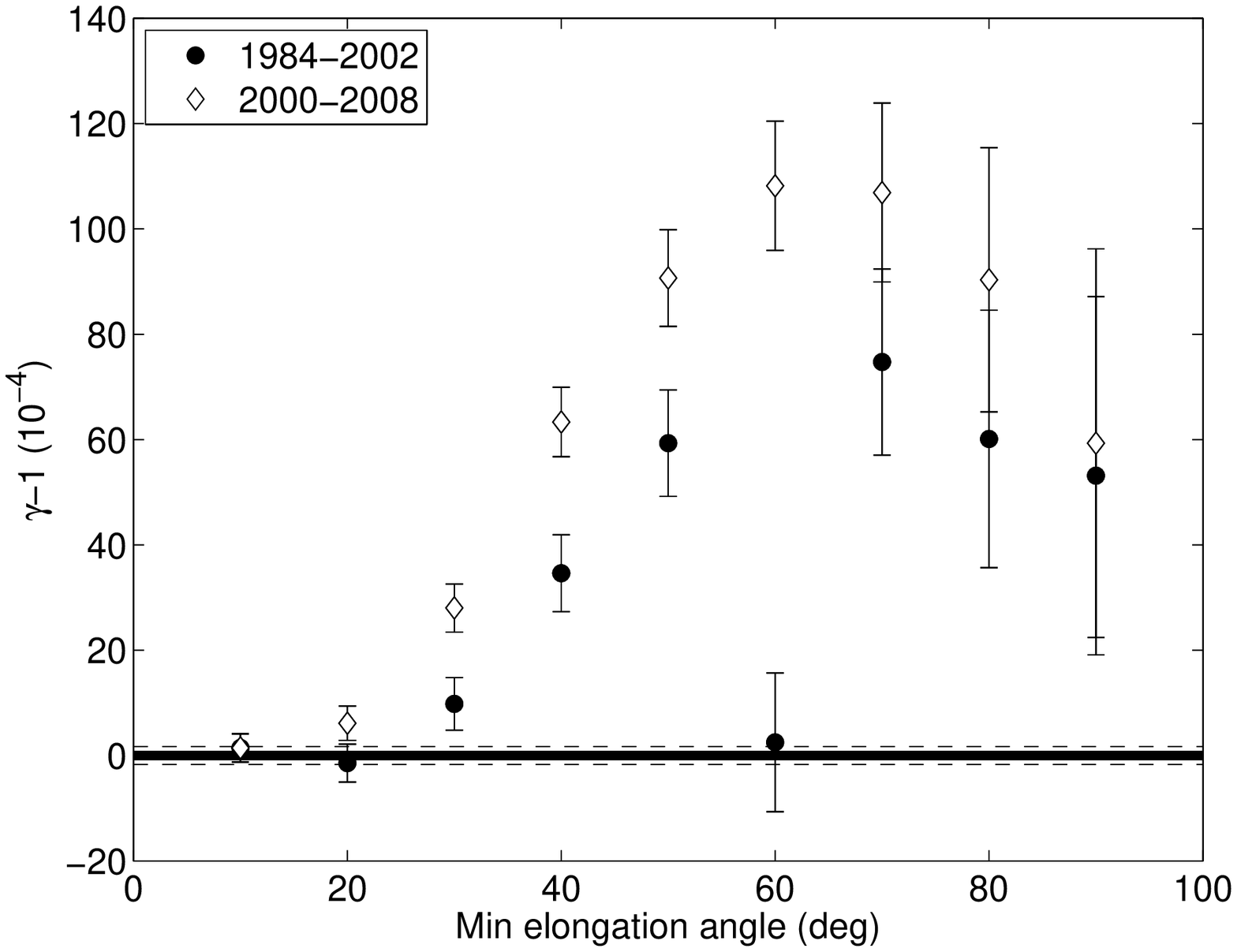}
\includegraphics[width=8cm]{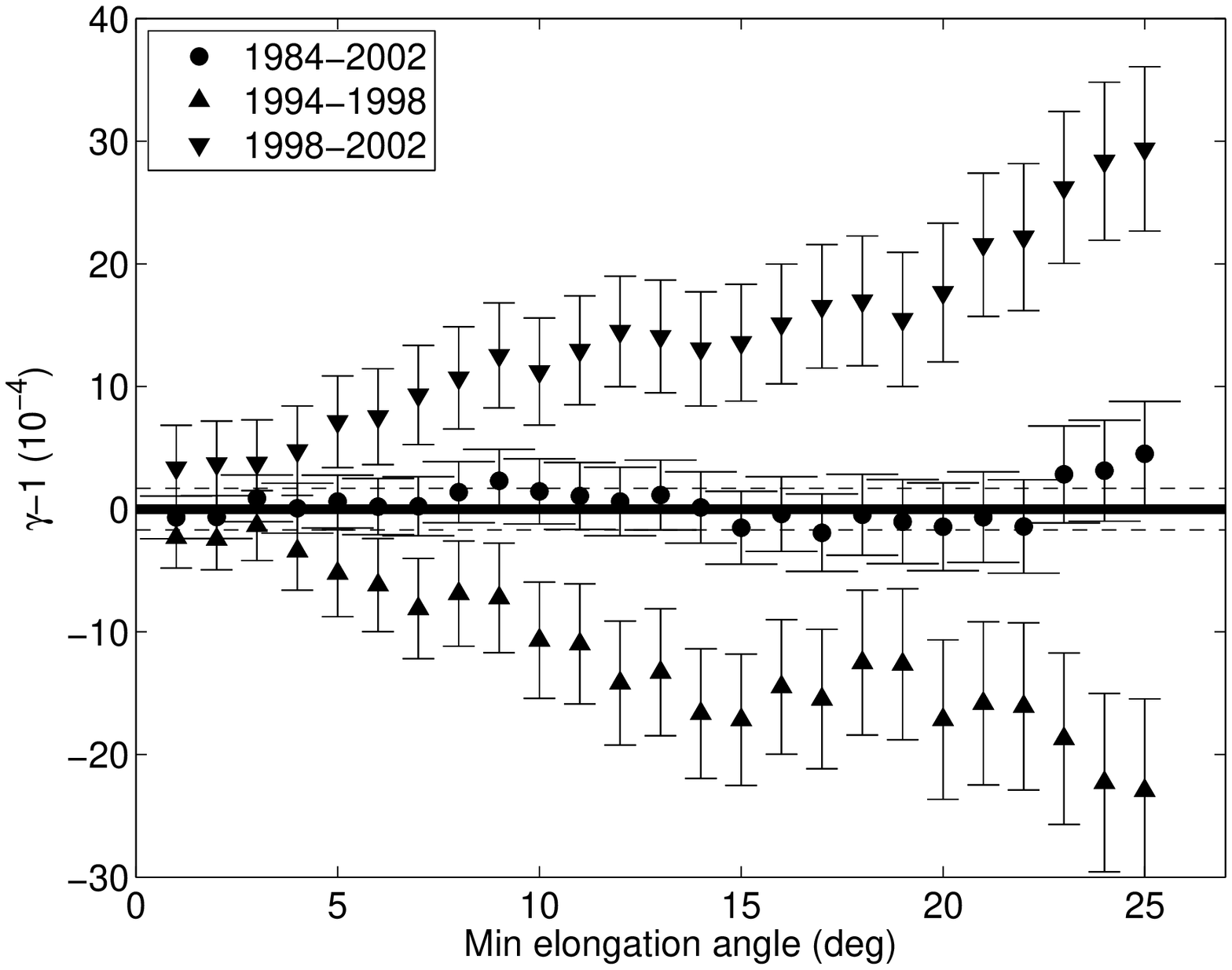}
\end{center}
\caption{Estimates of $\gamma$ for various cut-off of the Sun elongation angle. 
Horizontal, dashed lines figure $\pm\sigma$.}
\label{figelontest}
\end{figure*}

Since source coordinates are estimated during the analysis process, no
significant influence of the a priori catalogue is expected on $\gamma$.
To check this, we ran the previous solutions several times, using several
a priori catalogues. All of them were obtained after a global inversion 
of data over 1984--2008, wherein the celestial reference frame was maintained 
by applying a no-net rotation constraint on the coordinates on a well-chosen
subset of sources that defines the axes of the International Celestial 
Reference System (ICRS, Feissel \& Mignard~1997). Several subsets
achieve this goal (Ma et al.~1998, Feissel-Vernier~2003, Feissel-Vernier 
et al.~2006, Lambert \& Gontier~2009), and ensure an alignment of the output 
catalogue onto the ICRS within 0.05~mas. (The latter nevertheless decreases 
this value below 0.02~mas.) It finally appeared that the sensitivity of
estimated $\gamma$ to the chosen set of defining sources and to the a priori 
catalogue is at the level of $10^{-8}$, {\bf which is not statistically 
significant.}

We wondered whether the fit could be improved by removing data from
sources having a poor observational history (e.g., less than 2 observations
or observed in less than 3 sessions). We therefore ran one more time all
the above solutions after having downgraded about 200 sources as 
session parameters and suppressed the delays from another 100. Final 
post-fit root mean square (rms) and  normalized reduced $\chi^2$ per degree 
of freedom ($\chi^2/f$) were not changed significantly. (The $\chi^2/f$ is 
output by the VLBI analysis software and reflects the goodness of the fit
of the solution, including all adjusted parameters.) Influence on 
$\gamma$ estimates was only noticed at the level of 10$^{-6}$, which appears
to be non statistically significant, provided the standard errors reported
in Table~\ref{tab01}.
 
The post-fit rms delay of the solutions ranges 25--28~ps, 
Such a rms corresponds to a rough expected accuracy of 0.27~mas in 
terms of individual source positioning. One can readily see that, 
assuming such a measurement error on the direction of a grazing ray 
one can expect an error $\delta\gamma$ not lower than 
$\delta\gamma/\gamma\simeq\delta\theta/\theta\simeq1.5\times10^{-4}$. This 
is confirmed by the standard errors reported in Table~\ref{tab01}. 

\begin{figure*}[htbp]
\begin{center}
\includegraphics[width=8cm]{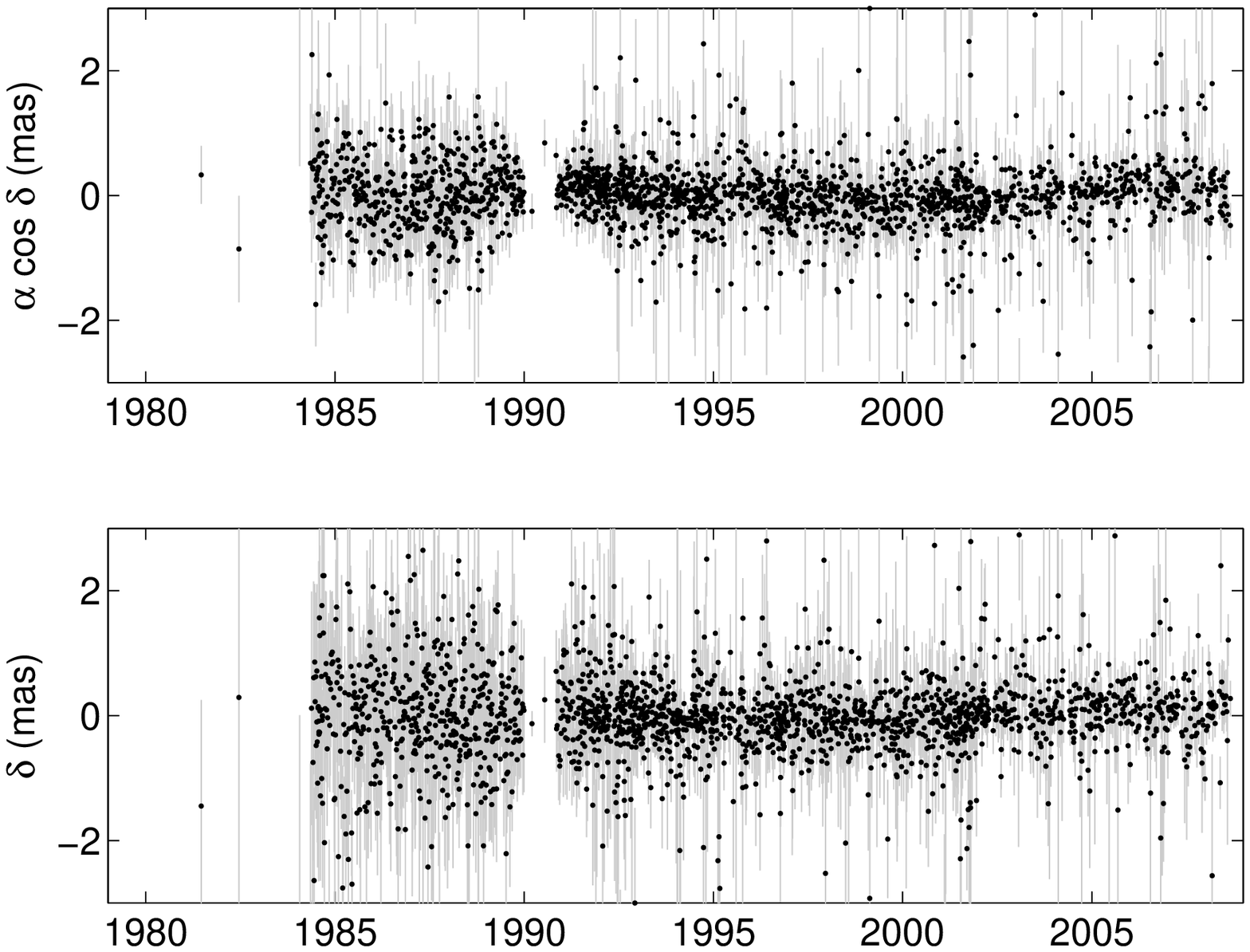}
\includegraphics[width=8cm]{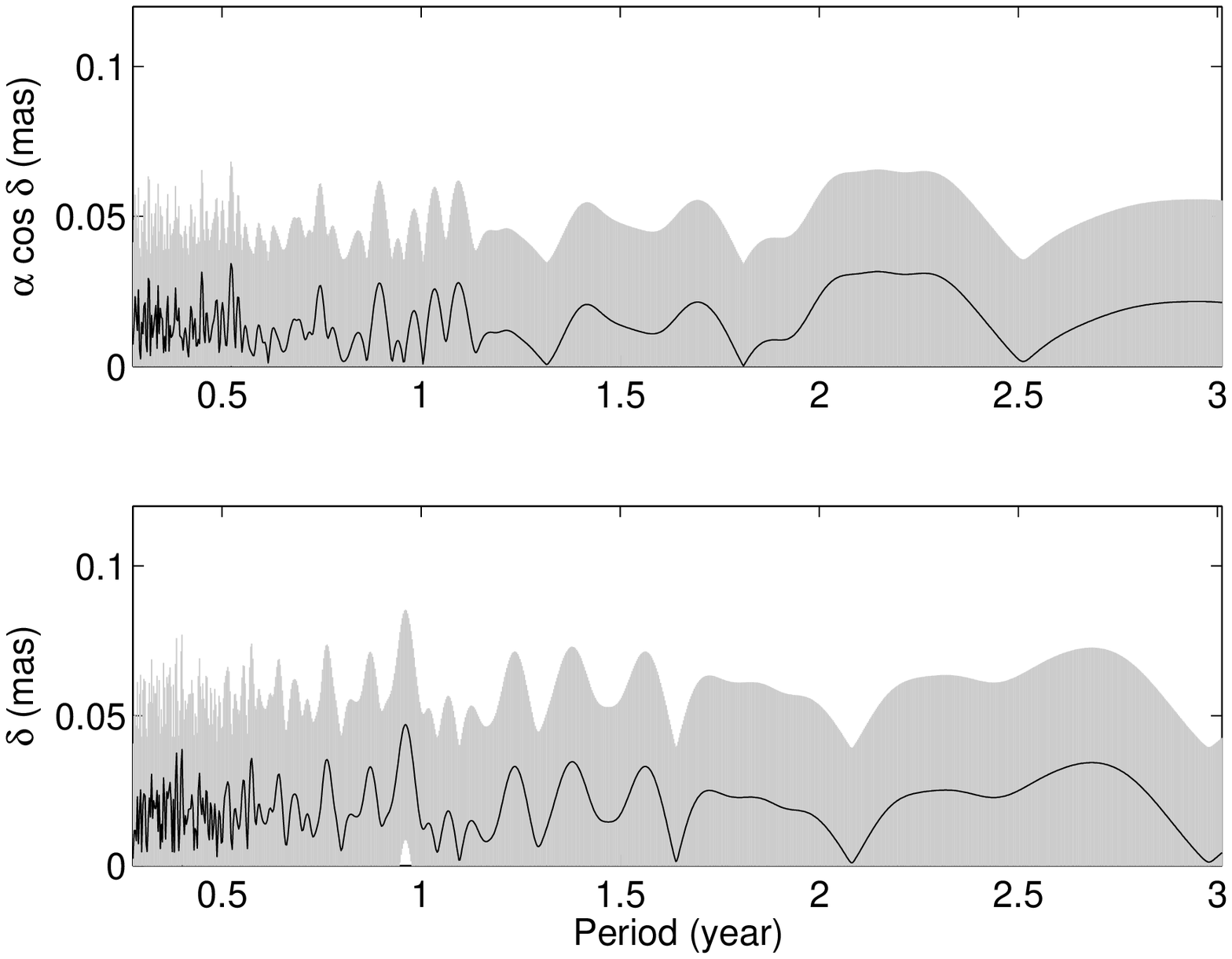}
\end{center}
\caption{{\it (left)} Session-wise coordinates of 0229+131. {\it (right)} 
Least-squares spectrum.}
\label{figtemps}
\end{figure*}

The solutions over 1979--1999, 1979--2008, 1984--2002 and 
1984--2008, that include a large number of sessions and delays, 
and that all have $\chi^2/f$ larger than 0.86, all result in estimates of 
$\gamma$ consistent with GR within $\sigma=2\times10^{-4}$. Using the 
sessions after 2002 or before 1979, {\bf that do not contain} close approaches 
below 15$^{\circ}$, makes the estimate of $\gamma$ depart from unity at 
the level of~$1\sigma$. Incidentally, the solution over 1979--1999 confirms 
the analysis of Shapiro et al. with a slightly lower formal error that may 
originate from a different analysis strategy and a different observational 
data set.

Although having a similar number of observations or sessions, solutions 
over 1998--2002(a,b) bring a standard error substantially higher than
1994--1998. Moreover, for 1994--1998, estimate of $\gamma$ appears to be 
lower than~1, whereas it is larger for 1998--2002(a,b). It indicates that 
the bending of sources is higher in the latter case. Intense solar 
activity during the latter period could be at the origin of the 
discrepancy: during periods of high activity, the higher electronic 
content results in a higher deflection of radio waves. The absence in the 
software of a specific modeling of solar plasma effects and the strong 
correlation of an uncorrected plasma-induced bending with the relativistic 
deflection prevent from separating these two phenomena.

\subsection{Dependence on the elongation angle}

To address the problem of the elongation angle to the Sun, we run 
several solutions with an increasing cut-off angle, removing sources below
successive thresholds up to 40$^{\circ}$. We apply this analysis scheme
over several time spans (Fig.~\ref{figelontest}). For 1984--2002, a 
substantial degradation of the estimates occurs beyond 25$^{\circ}$, 
in agreement with similar tests in Shapiro et al. A bump reaches a
maximum around 60$^{\circ}$ and then, estimates of $\gamma$ get closer
to unity. We run a similar analysis over 2000--2008 because it
constitutes a data set decorrelated from the one used by Shapiro et al.
(allowing for the fact that (i) a part of the observed sources and observing
antennas are the same in both data sets, (ii) the latter contains substantially
less sessions than the former). The bump also shows up using this data set.

We also checked what happens at short elongation angles over 1984--2002, 
1994--1998, and 1998--2002. Below 25$^{\circ}$, the deviation from unity 
stays within the error bars with non statistically significant variations. 
For shorter solutions, estimates rapidly degrade beyond an elongation 
cut-off of a few degrees. For 1994--1998 and 1998--2002, the 
degradation occurs in opposite directions. Estimates of $\gamma$ 
{\bf appear} to be lower than 1 in the former case, while
they are larger in the latter, consistently with the global
estimates of $\gamma$ over the same time periods shown in Table~\ref{tab01}.
The possible reason of such differences has already been addressed.

\subsection{Approach based on radio source coordinate time series} \label{sects}

Estimating session-wise coordinates of sources can also be a mean 
of looking at a possible deflection when the sources travel in the vicinity
of the Sun. An uncorrected bending should appear as an annual signal in 
coordinate time series.

\begin{figure}[htbp]
\begin{center}
\includegraphics[width=8cm]{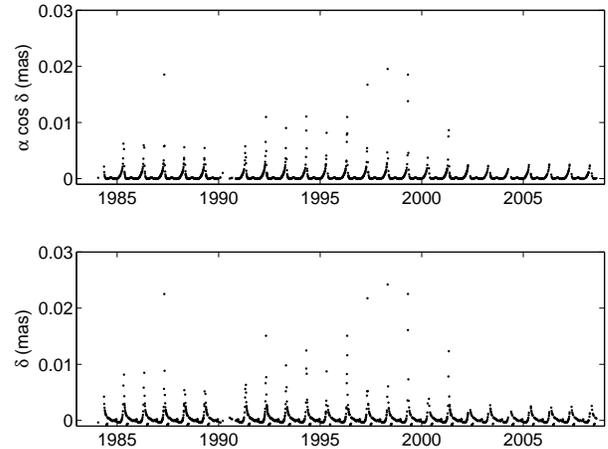}
\end{center}
\caption{Theoretical extra deflection from GR model for
 0229+131 for $\gamma-1=2\times10^{-4}$.}
\label{figgr}
\end{figure}

Among the observed sources, only two have close approaches below
2$^{\circ}$ and are observed in more than 500 sessions. Both cases
are similar, and we will only {\bf treat the source that has the largest
observational history:} 
0229+131 (quasar 4C~13.14). We obtained a coordinate time series 
using the analysis strategy of section \ref{secglo}, except that 
$\gamma$  is now fixed to 1, and coordinates of 0229+131 are 
estimated per session. The closest approach to the Sun is 
$\sim$1.5$^{\circ}$. At that time, the expected deviation, 
following Eq.~(1), is $\sim$0.3$^{\prime\prime}$. Parameter 
$\gamma$ being fixed to unity, 
this deflection is already corrected and will not show up in the 
coordinate time series. Obtained right ascension and declination 
time series are displayed in Fig.~\ref{figtemps}. The spectrum does 
not show any significant peak at annual period, indicating that no
extra deflection is detectable. Assuming an hypothetic deviation 
of $\gamma-1$ of $2\times10^{-4}$, the incremental deflection 
would be as drawn in Fig.~\ref{figgr}. Peaking at $\sim$0.03~mas, 
it is therefore not detectable in the spectrum. It follows that 
examination of coordinate time series for 0229+131 can only 
constrain $\gamma$ to be close to unity at approximately 
the same level of accuracy already obtained from global estimates.

Note that evolution in source structure can show up in coordinate time 
series at lower frequencies, as medium or long-term patterns (a few months 
to years), like the slight curvature showing up in right ascension 
{\bf plotted in Fig.~\ref{figtemps}.
About relations between source structure and coordinate time series, the
reader can refer to, e.g., Fey et al.~1997, that treats the case
of the quasar 4C~39.25.}

\section{Discussion and conclusion}

In the above sections, we have used several methods to look for radio wave
deflection in the vicinity of the Sun, starting from a 30-yr long routine 
geodetic VLBI observational data base. We interpret this deflection 
in terms of gravitational bending, as expressed in Eq.~(1). Using 
several strategies and various data sets covering different time spans, 
we arrive at the conclusion that $\gamma$ is unity within $2\times10^{-4}$. 
The estimate of $\gamma$ can even reach values close to 
unity by $7\times10^{-5}$ when the time span is limited to 1984--2002,
i.e., to sessions containing observations of sources at less than 15$^{\circ}$ 
to the Sun. Using longer time spans, although decreasing the formal error 
due to a larger number of observations, makes the estimates depart from 
unity by about~$1\sigma$.

The main factor of limitation is the uncertainty in the determination of 
(global or session-wise) radio source coordinates.
Causes of this uncertainty {\bf have} been addressed in various works 
(see, e.g., Ma et al.~1998, Gontier et al.~2001). The 
VLBI-derived apparent position of a source may change with the 
global orientation and shape of the antenna array when the structure
of the source is extended or not circular. Fey \& Charlot (1997), 
using Very Long Baseline Array (VLBA) maps at 2 and 8~GHz, provided estimates 
of the structure delay arising from the extended character of the source. 
In our example of section \ref{sects}, the structure of 0229+131 is 
expected to bring an extra delay below 3~ps, let 0.03~mas 
(see also Ma \& Feissel~1997), that partially explains the noise 
level observed in Fig.~\ref{figtemps}. It turns out that, in absence 
of a direct correction of the delay, based on, e.g., instantaneous maps 
of the source, {\bf the accuracy of $\gamma$ estimates from time series 
analysis cannot be better than 10$^{-4}$.}

Other potential sources of error are the mismodeling of the propagation 
delay through the troposphere, as well as deficiencies 
in the network (e.g., change of  geometry and performances from one 
session to another, dissymmetry between North and South hemispheres). 
The amplitude of the noise that emerges from them remains difficult 
{\bf to be precisely quantified} at this time. It is generally admitted that 
it is as large as the effect of source structure.

Note also that derivation of radio source coordinate time series implies
a robust maintenance of the celestial and terrestrial reference frames,
so that frame effects do not introduce spurious perturbations of the 
estimated coordinates. During the derivation process, we checked
various analysis strategies and we noticed that, when the celestial
frame is not sufficiently maintained (e.g., when too few sources are
constrained by the NNR), a semi-annual peak could appear at $3\sigma$.
In a similar way, fixing the station coordinates to their ITRF values
introduces an annual term at the same level. These spurious peaks, that
could lead to erroneous physical interpretations in the present context,
are good illustrations of the sensitivity of VLBI to reference frames.

Already mentioned is the mismodeling of the solar corona contribution to 
light scattering and bending. Although this source of error is
neglected for geodetic purposes when radio sources are observed at large
elongations to the Sun, it becomes crippling for tests of GR since 
observers do need to observe as close as possible to the Sun. From 
section \ref{secglo}, we tend to conclude that fluctuations in solar 
coronal plasma limit the accuracy of $\gamma$ estimates at the same 
level of the above-listed sources of error.

Thus, various instrumental and modeling errors and analysis strategy defects, 
that cannot be decorrelated and corrected yet, explain the current 
limitation of VLBI for estimating~$\gamma$.

Compared to the error reported in Shapiro et al., we do not consider that 
we have substantially improved the determination of $\gamma$. 
The slight gain in accuracy can be attributed to the 
extra years of data (1999--2008) of which the first 4 years (1999--2002) 
are rich in close approaches, along with the improvement of the quality 
of the VLBI network and observations during this time. Our work nevertheless 
constitutes {\bf an independent check}, and provides some qualitative insights 
into systematics that show up in the analyses of the current geodetic VLBI 
observational database.

To conclude, we wish to mention that, although current VLBI appears to be 
not competitive with spacecraft systems for relativistic experiments, the 
huge number of VLBI measurements, in all directions and at a large number
of epochs, constitues an interesting potential for testing other 
theories than the PPN formalism, as for example the scenario of 
Jaeckel \& Reynaud~(2006) where parameter gamma is replaced by a 
function depending on the elongation angle.

\begin{acknowledgements}
We are grateful to Drs. Anne-Marie Gontier and Peter Wolf (Observatoire de Paris)
for useful discussions about possible tests. We thank Prof. Harald Schuh for his
review that helped in improving the paper. This study could not have 
been carried out without the work of the International VLBI Service 
for Geodesy and Astrometry (IVS) community that coordinates observations, 
correlates and stores geodetic VLBI data.
\end{acknowledgements}

\end{document}